# Self-Assembly of Semiconducting Single-Walled Carbon Nanotubes into Dense and Aligned Rafts

Justin Wu[1†], Liying Jiao[1, 2†], Alexander Antaris[1], Charina L. Choi[1], Liming Xie[1], Yingpeng Wu[3], Shuo Diao[1], Changxin Chen[1], Yongsheng Chen[3] and Hongjie Dai[1]*

1. Department of Chemistry and Laboratory for Advanced Materials, Stanford University, Stanford, California 94305, USA

2. Key Lab of Organic Optoelectronics & Molecular Engineering, Department of Chemistry, Tsinghua University, Beijing 100084, China

3. Key Laboratory of Functional Polymer Materials and Centre for Nanoscale Science and Technology, Institute of Polymer Chemistry, College of Chemistry, Nankai University, 300071, Tianjin, China

*†* These authors contributed equally to this work

∗ Correspondence to hdai@stanford.edu

**Single-walled carbon nanotubes (SWNTs) are promising nanoelectronic materials[1-3] but face long-standing challenges including production of pure semiconducting SWNTs and integration into ordered structures. Here, highly pure semiconducting single-walled carbon nanotubes are separated from bulk materials and self-assembled into densely aligned rafts driven by depletion attraction forces. Microscopy and spectroscopy revealed a high degree of alignment and a high packing density of ~100 tubes/µm within SWNT rafts. Field-effect transistors (FETs) made from aligned SWNT rafts afforded short channel (~150 nm long) devices comprised of tens of purely semiconducting SWNTs derived from chemical separation within a < 1 µm channel width, achieving unprecedented high on-currents (up to ~120 µA per device) with high on/off ratios. The average on-current was ~ 3-4 µA per tube. The results**

**demonstrated densely aligned high quality semiconducting SWNTs for integration into high performance nanoelectronics.**

One approach to address the challenges of carbon nanotube electronics is to combine controlled growth of aligned SWNT arrays on a surface with removal of metallic SWNTs (m-SWNTs) during or after growth.[4-7] The process of utilizing crystal steps on annealed quartz[5,7] or sapphire[6,8] and/or ultralow gas flow[9] to grow SWNTs by chemical vapor deposition (CVD) into well-aligned arrays has been demonstrated. There are several approaches to selectively remove metallic tubes from the aligned arrays, including electrical break-down,[5] plasma etching,[10] ultraviolet irradiation,[7] functionalized Scotch tape removal,[6] and so on.[11] Most of these approaches can successfully remove m-SWNTs, but at the expense of partial s-SWNT destruction. The density of nanotubes in these arrays is limited by CVD growth (<15 tubes/μm)[12] and the number of transfers (up to five with decreasing efficiency),[13,14] and is further decreased by the removal of metallic tubes. In addition to this 'dry' strategy, there are several wet approaches for the solution-phase separation of m- and s-SWNTs, including density gradient ultracentrifugation (DGU),[15,16] dielectrophoresis[17] and chromatography techniques.[18-21] These methods can produce highly pure s-SWNTs and even s-SWNTs of specific chirality.[19,20] It remains challenging to control the arrangement of those sorted SWNTs when placing them from a solution onto substrates for device integration. Dielectrophoresis has been utilized for dense assembly of SWNTs between electrodes; however, this method has been limited to 25 tubes/μm, with a possible preference for residual metallic tubes.[22] Self-assembly is a unique 'bottom-up' approach for making well-ordered architectures using various nanomaterials as building blocks.[23] It has been shown that a self assembly approach can place SWNTs into patterned trenches with controlled trench density, but the inter-tube relations (density and alignment) within a trench are not defined or controlled.[24] The Langmuir Blodgett and Langmuir Schaeffer method have been used to densely align SWNTs,[25] although FETs fabricated on the arrays show less than 0.2μA of current per SWNT.[26] In this work, we developed a method for densely packing enriched s-SWNTs into arrays through the self-assembly of well-aligned raft-like structures. Semiconducting SWNTs from raw material synthesized by the arc discharge process[27] were extracted by a single step gel filtration or density gradient ultracentrifuge and then assembled to produce aligned semiconducting rafts on substrates. We found that s-SWNTs dispersed in sodium cholate (SC) and sodium dodecyl sulfate (SDS) solutions could close-pack into flat rafts with a typical raft length of > 5 μm and width of > 1 μm, a process driven by depletion attraction during water evaporation on a (3-aminopropyl) triethoxysilane (APTES) modified $SiO_2$/Si substrate. The rafts were characterized by atomic force microscopy (AFM) and Raman spectroscopy, and were also made into field effect transistors (FETs). The high purity and high quality of densely aligned SWNTs (~100 tubes/μm) led to the highest performance short-channel multi-tube semiconducting FETs based on chemically sorted semiconducting SWNTs to date in both total current per device and current density.

To obtain semiconducting-enriched SWNTs, we separated SC- and SDS-coated, stably suspended arc-grown SWNTs by modifying the gel filtration technique[21,28-30] using an allyl dextran-based size-exclusion gel (Sephacryl S-200, GE Healthcare). The metal/semiconductor selectivity of this separation approach originates from distinct interactions between different types of SWNTs, surfactants, and gel[20,21]. Metallic SWNTs in solution are more reactive and fully surfactant-coated than their semiconducting counterparts and therefore interact more weakly with the gel; these tubes were eluted first from the column. We used a modified gel filtration procedure to reach high selectivity through a single filtration by washing with several eluent solutions composed of increasing SC to SDS ratios to increase the 'dispersing power'[31] for eluting metallic species first (see Supporting Information for details). The third wash with a 0.75 wt% SC/0.25 wt% SDS solution eluted a dark red fraction (S-fraction), indicating a high concentration of s-enriched SWNTs (**Fig. 1c**).

Atomic force microscopy (AFM) was used to investigate the length distribution of the s-SWNTs (Fig. S1). The mean SWNT length was ~ 0.7 μm, with SWNT lengths up to ~1.8 μm. We characterized the as made and sorted fractions of SWNTs by Raman and UV-*vis*-NIR absorption spectroscopies in solution. Raman spectra of the as-made solution and s-SWNT fraction, mass-normalized by the π→π* interband transition ~4.5 eV,[32,33] were measured using both 532 nm and 785 nm laser excitation (Fig. 1a-b). The radial breathing mode (RBM) at ~170 cm$^{-1}$ corresponds to s-SWNTs with a diameter of ~1.5 nm in resonance with 532 nm laser excitation, while the peak at ~160 cm$^{-1}$ is due to m-SWNTs (diameter ~1.6 nm) in resonance with 785 nm laser excitation (see Supplementary Information for the assignment of s- and m-SWNTs and diameter estimation). As expected, the S-fraction showed a stronger semiconducting peak (Fig.1a) and highly diminished metallic peak (Fig.1b) compared to the unseparated solution. The intensity ratios of the 785 nm-excited m-SWNT peak to the 532 nm-excited s-SWNT peak were used to estimate an s-SWNT purity of ~97% in the separated solution using our previously reported method.[29] UV-*vis*-NIR absorption spectra of the S-fraction demonstrated a large decrease in the intensity of metallic $M_{11}$ peak near 700 nm (Fig.1c), further indicating the removal of m-SWNTs by gel filtration. With our single-step gel-filtration method, we obtained a large volume of s-SWNTs in solution. The relatively large diameter (~1.2-1.7 nm) of s-SWNTs obtained by this approach makes them more suitable for producing electronic devices with high mobility and high on-state currents than smaller diameter HiPco SWNTs.[34]

The next step towards device integration of the sorted s-SWNTs is to assemble them into aligned or other ordered structures. We found that s-SWNTs could be self aligned into raft assemblies by utilizing depletion attraction forces originated from entropic effects at high surfactant loading.[35-37] To achieve self-assembly of s-SWNTs (**Fig. 2a**), we first placed a drop of s-SWNT solution, in which the total SC and SDS surfactant concentration was above its critical micelle concentration (CMC), on an APTES-modified $SiO_2$/Si substrate. In order to judge the concentration of tubes within the solution, a UV-*vis* spectrum was taken, revealing an optical density (OD) of 0.31 at the π→π* transition ~278 nm. We then let water partially evaporate from the droplet until 50% of the volume remained before drying the substrate with a directional gas flow.

AFM showed that s-SWNT rafts with sizes of up to a few square micrometers could be made reliably under the optimized conditions (Fig.2b-e). The density of s-SWNTs within these rafts reached up to ~100 tubes/µm. The thickness of the raft was 1-2 times of the diameter of s-SWNTs as measured by AFM, indicating mostly a mono-layer assembly. Figures 3 a-d show an AFM image and RBM Raman mapping images of the same s-SWNT raft measured under 532, 633 and 785 nm excitations, respectively. Despite the stronger resonance of m-SWNTs at 785 nm excitation compared with s-SWNTs at 532 nm excitation (Fig. 1a-b), a much lower signal from m-SWNTs (RBM~198 cm$^{-1}$ @ 633 nm, RBM~160 cm$^{-1}$ @ 785 nm) than s-SWNTs (RBM ~153 and ~171 cm$^{-1}$ @ 633 nm, RBM~170 cm$^{-1}$ @ 532 nm) (Fig.S3) was observed, demonstrating substantial s-SWNT enrichment in the rafts. In polarized Raman spectroscopy measurements, the semiconducting RBM modes (**Fig. 3**e-g) from the same raft in Fig. 3a showed the maximum intensity along the same direction of the raft alignment direction, indicating good alignment of s-SWNTs along the raft direction.[25] No obvious Raman D-band was observed in these s-SWNTs rafts, demonstrating that the separation and assembly procedures retain the high quality of SWNTs.

We attribute self-assembly of SWNTs into aligned rafts to depletion attraction forces, previously observed to achieve self-assembly of cellular macromolecules,[38] colloidal nanorods,[39] and silver nanocrystals.[40] Briefly, a system containing nano-objects like carbon nanotubes and many depletion agents such as surfactant micelles, obtains maximum entropy when a maximum volume is available to the depletion agents.[41] A so called depletion zone exists around the shell of each object that excludes the agents. When two objects approach each other during solvent evaporation, the excluded volumes (or depletion zone) around the objects begin to overlap, giving an increase in the volume accessible to the depletion agents (e.g., surfactant micelles in the current case) and thus increase the entropy of the entire system. This provides an attractive force between the objects and promotes alignment of rod- or tube-like objects to maximize the accessible volume. In our case, the depletion attraction-driven self-assembly of s-SWNTs at the solid-liquid interface is clearly affected by surfactant composition and concentration, substrate functionalization, and degree of solvent evaporation. We systematically investigated the effects of these three factors on raft formation to achieve optimized and densely packed s-SWNT arrays.

We first studied the role of surfactant composition in the formation of SWNT raft arrays. Surfactants act in two important ways in the self-assembly process: first, as depletion agents in solution, and second, as surface ligands to stabilize SWNT suspension. The competition and balance of these two factors and tube-substrate interactions at the APTES coated $SiO_2$ substrate led to SWNTs assembly. We determined the optimized surfactant conditions for raft formation using s-SWNTs. The composition of surfactant in the freshly separated S-fraction was 0.25 wt% SDS and 0.75 wt% SC, slightly above the CMC of SDS (~0.24 wt%) and much higher than the CMC of SC (~0.6 wt%). Under this condition, the s-SWNTs are stably suspended in solution due to coating by a high percentage of SC molecules. The s-SWNTs coexist with SC and SDS micelles in this solution. With this solution, we found individually distributed SWNTs on the substrate as evidenced by AFM imaging (see Supplementary Information, Fig S2a), indicating a weak depletion force and stable coating against SWNT packing at this surfactant composition. As the percentage of SDS increased, the range of the depletion force between SWNTs increased due to the much larger excluded volume of SWNTs to SDS micelles caused by the size difference between SDS micelles (aggregation number $N_A$=64)[42] and SC micelles ($N_A$=4).[43] Additionally, SDS is a poorer coating surfactant than SC, resulting in less stable SWNT suspensions. Upon increasing the SDS concentration to 1 wt% with no added SC, SWNTs formed parallel closely packed arrays with some big bundles (Fig. S2b), indicating a large depletion force and low stability of SWNT suspension.

Based on these results, we tuned the conditions for self-assembly by mixing varying amounts of SC and SDS to suspend SWNTs while maintaining the total concentration of SDS and SC at 1 wt%. We found the optimized surfactant composition to assemble s-SWNTs into raft arrays to be ~0.9 wt% SDS and ~0.1% wt% SC (Fig. 2b-e), achieving a balance between depletion force and stability of surfactant coating on SWNTs against packing. When deviating from this condition, higher percentages of SDS resulted in thicker and narrower bundles whereas lower percentages resulted in fewer and narrower rafts. Assembly results obtained at other surfactant compositions are further discussed in Supplementary Information (Fig. S2c-f).

Substrate surface modification also affected the formation of s-SWNTs rafts. We found that APTES modification of $SiO_2$ improved the deposition of SWNTs on $SiO_2$/Si substrates, likely due to attractive interactions between the amine groups and surfactant coated SWNTs.[44] Depletion attraction between SWNTs and the substrate additionally aids deposition. The substrate-SWNT interactions facilitated the formation of 2D rafts instead of 3D bundles during solvent evaporation, as s-SWNTs spread over the planar substrate to maximize attractive interactions with the substrate. Without APTES functionalization, very few s-SWNTs were deposited on the substrate (Fig. S3a).

The degree of water evaporation is another factor in successful assembly. Depletion attraction is a short-range force and therefore, the distance between tubes needs to be decreased before tube-tube attraction can play a role. To confirm this, we decreased the rate of water evaporation by sealing a $SiO_2$/Si substrate with a drop of s-SWNTs solution in a humidified container with water (inset of Fig. S3b). After 24 h, no obvious shrinkage of the droplet was observed. After blow drying, we only observed mostly aligned single tubes or small bundles on the substrates (Fig.S3b), suggesting the necessity of water evaporation to bring SWNTs into the effective depletion distance. Water evaporation also increases the surfactant concentration, associated with the increase of micelle density. Too much or too rapid evaporation resulted in thick assemblies and large aggregations (Fig. S3c-f). Directional blow-drying after the evaporation helped to orient individual rafts approximately along one direction on the substrates (Fig. S3g and h).

The self-assembly technique could be easily generalized to solution sorted SWNTs by other sorting methods. We succeeded in making SWNT rafts from solutions of arc-discharge grown 'P2 SWNTs' (Carbon Solutions, Inc) sorted by DGU[15] (see Supplementary Information for experimental details of the DGU). We exploited the high quality, purity, relatively large diameter, high packing density and good alignment of s-SWNTs for building field effect transistor (FET) devices. With the DGU sorted semiconducting SWNTs, we fabricated FETs with short (~150 nm) channel lengths on rafts annealed in vacuum at 700 °C using 20-nm-thick Pd as source and drain electrodes, 10 nm thick $SiO_2$ as dielectric, and $p^{++}$ Si as the back gate. The electrodes were fabricated using standard electron-beam lithography and the width of the electrodes were 180 nm (which set the contact length of source and drain electrodes with SWNTs to be ≤ 180 nm) as imaged by AFM (Figure 4c, inset). To improve the contacts, the devices were annealed in vacuum at 200 °C for 20 min before electrical measurements in ambient.[45]

The highest on current achieved by the semiconducting raft FET devices on 10 nm $SiO_2$ was 121 µA under a 1V bias within an effective channel width of ~ 1 micron (Fig. 4b,c), suggesting a high current density of 125 µA/µm. Effective channel width was measured as the distance from the leftmost SWNT to the rightmost SWNT contained within the device (Fig. 4a-c). The on/off ratio of this device was ~ 4000, again, measured at 1V bias. The number of SWNTs within this device was counted to be ~ 37, corresponding to an average of 3.2 µA of current per SWNT. The highest current density achieved was 160 µA/µm in a different device with a raft width of ~500 nm containing ~ 39 SWNTs, affording a current of 80 µA. Multiple devices were found to have current on this order while maintaining a high on/off ratio larger than $10^3$ (Figure 4d). The average current per SWNT over the device set was found to be 3.0 µA per SWNT, matching devices that had been made on individual SWNTs from the DGU sorted P2 solution (see Supplementary Information).

Our devices compare favorably with several reports from literature. Based on our relatively high current per SWNT, our process and device structure maintain the quality of the SWNT and minimize the contact resistance degradation associated with dense SWNT proximity.[26] The ~ 3 µA on-current per SWNT was the highest for high density (> 10/µm) multi-tube FETs with on/off > 1000.[24,29,47-50] It was the first time that >100 µA total on-current was reached for SWNT short channel FETs with on/off

>1000 using sorted CNTs in a sub micron parallel short channel. Our highest measured current density of 160 µA/µm exceeds the highest previously reported (125 µA/µm[26]) for semiconducting CNT devices. Recent work in this field showed chemical assembly of sorted semiconducting SWNTs into well ordered channels of increasingly small pitch down to 100 nm.[24] However, at this pitch, the achievable density of CNTs within a single small device is still very limited due to a CNT/channel number of < 3 in a channel width of 70nm at a pitch of 200nm, limiting the current density of parallel CNT devices to 20 µA/µm. Our method of packing CNTs together allows for much closer proximity of CNTs and consequently higher currents. On the other hand, a method for even denser packing of CNTs over large areas at densities of >500/µm using the Langmuir Schaeffer method has also been published recently.[26] Devices fabricated on SWNTs at this density show very high contact resistance due to the close proximity of SWNTs, with current per SWNT on average < 0.2 µA. This limits on-current density to a comparable value of 125 µA/µm despite their higher SWNT/µm density. In addition, a lower current per SWNT requires substantially higher purity to achieve the same on current without m-SWNTs reducing the on/off ratio. They were limited to 100nm wide channels carrying only 12.5 µA of total current. In contrast, our devices show much higher current per SWNT, which allows for a higher current density in our FETs.

In summary, we separated high purity semiconducting SWNTs by a new gel filtration method and assembled them into flat, densely-packed and aligned arrays on $SiO_2$/Si substrates utilizing depletion attraction between SWNTs and attractions with the substrate. We investigated the effects of the composition of surfactant, substrate surface modification and the evaporation of water on the formation of s-SWNT rafts. Self-assembly of s-SWNTs into dense 2D raft arrays up to 10 µm$^2$ and high density of ~100 tubes/µm were obtained reliably. Short channel FETs made on these rafts showed high performance, with on currents reaching over a hundred µA. Measurements showed high current density within devices containing only semiconducting nanotubes. The performance of raft electronic devices can be further improved by increasing the purity of s-SWNTs by further refining gel filtration or density gradient ultracentrifuge. Although raft positional control has not yet been achieved, chemical patterning of APTES on substrates could be one possible route to achieving this, further facilitating electronic applications. This simple, reliable and nondestructive assembly method opens up a new way to produce densely packed semiconducting SWNTs aligned arrays for high performance electronic devices.

**Acknowledgements**

This work was supported by Samsung (sorting and assembly), MARCO MSD, and U.S. Department of Energy, Office of Basic Energy Sciences, Division of Materials Sciences and Engineering under Award # DOE DE-SC0008684 (nanoscale alignment, nanofabrication and Raman instrumentation). C.L.C. was supported by the National Science Foundation under Award No. CHE-1137395.

**Figure captions**

**Figure 1.** Separation for enrichment of semiconducting single-walled carbon nanotubes. (a) and (b) Raman spectra of mass-normalized as made and s-SWNTs solutions under 532 nm excitation and 785 nm excitation, respectively. Raman measurement parameters were maintained for the same excitation wavelength. The purity of the semiconducting SWNTs may be estimated from the ratios of the s-SWNT RBM peak intensities under 532 nm excitation to the m-SWNT RBM peak intensities under 785 nm excitation (see Supplementary Information). (c) UV-*vis*-NIR absorption spectra of mass-normalized as made (black) and semiconducting-enriched (red) SWNTs solutions. Left inset, spectrum showing equal absorption intensity at the π→π* transition ~4.5 eV (273 nm)[29,30]. Right inset, photographs of as made (left) and S-fraction (right) SWNT solutions.

**Figure 2.** Formation of rafts of semiconducting carbon nanotubes. (a) Schematic of the process of assembly of s-SWNTs into aligned rafts driven by depletion force and gas flow. At a high loading of surfactant (significantly higher than CMC), numerous micelles coexisted with SWNTs. Entropic effects resulted in short-range attractive depletion force between SWNTs. Attractive forces also exist between SWNTs and amine functionalized substrates. These forces lead to SWNTs assembling into densely packed structures at the interface of water-substrate to increase the free volume available to micelles. Gas flow was then used to orient the 2D structures in an ordered way on the surface. (b)-(e) AFM images of semiconducting SWNT rafts.

**Figure 3.** Characterization of the semiconducting enriched SWNT rafts with Raman spectroscopy. (a) AFM image of a s-SWNT raft. (b-d) Raman mapping of the s-SWNT raft in panel (a) at 532, 633 and 785 nm excitations, respectively. The blue and red colors show the regions with semiconducting and metallic RBM signals, respectively. (e) Spectra of RBM peaks taken with 532 nm laser excitation at polarizations 0º to 90º relative to the raft orientation in 30º intervals. (f-h) Polar plot of RBM intensities after background subtraction to the laser polarization direction for different RBM shifts and excitations (170 $cm^{-1}$ using 532 nm, 171 $cm^{-1}$ using 633 nm, and 160 $cm^{-1}$ using 785 nm). Intensity scales are linear from zero intensity with arbitrary units. A $\cos^4(\theta)$ fitting to the intensity is shown in red.

Figure 4. Electrical devices formed on densely packed semiconductor enriched SWNT rafts. (a) Schematic of short channel device on 10 nm $SiO_2$. (b) Source-drain current vs. back gate voltage ($I_d$-$V_{gs}$) characteristics of FET made on s-SWNTs raft with a channel length of 140nm. Bias voltages: 1 V, 500 mV, 100 mV and 1mV, from top to bottom. (c) $I_d$-$V_{ds}$ curves of the same FET as in (a). $V_g$ was swept from -5 V to 0 V in steps of 0.5 V. Inset: AFM image of the device. (d) $I_d$-$V_{gs}$ characteristics of the next best five FETs made on s-SWNTs rafts. Colors correspond to the bias voltages of 1 V, 500 mV, 100 mV and 1mV, from top to bottom, blue, green, red, and black.

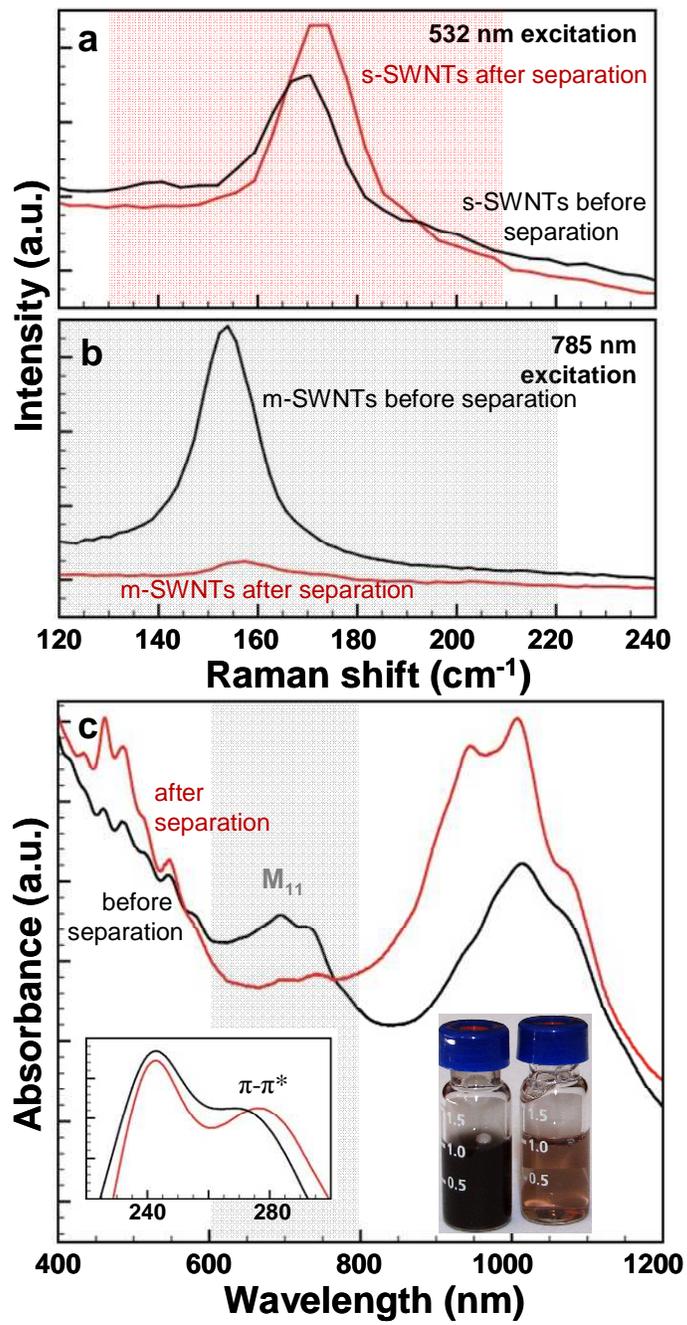

**Figure 1**

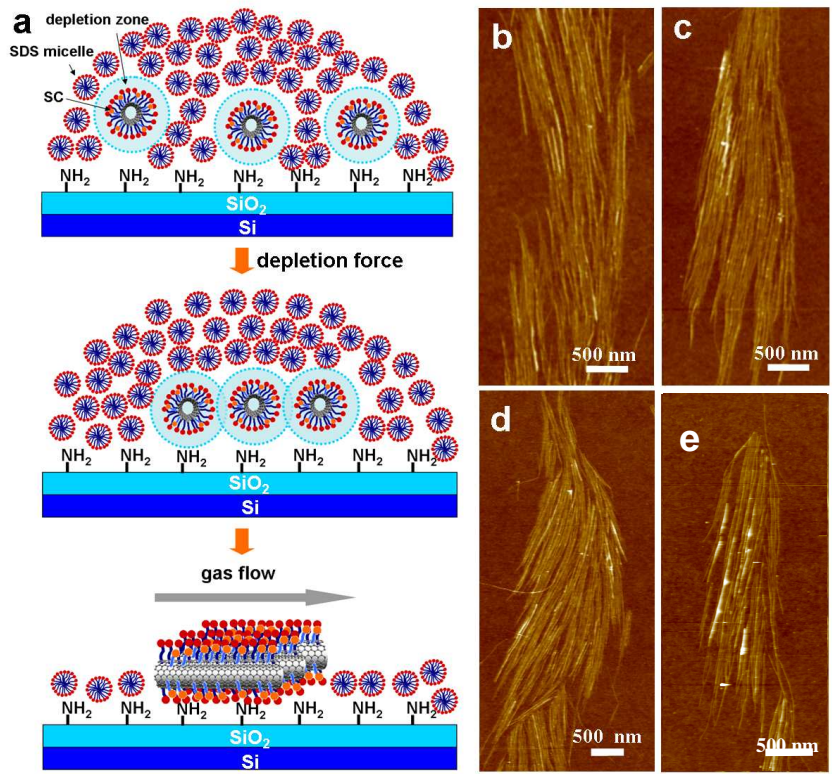

**Figure 2**

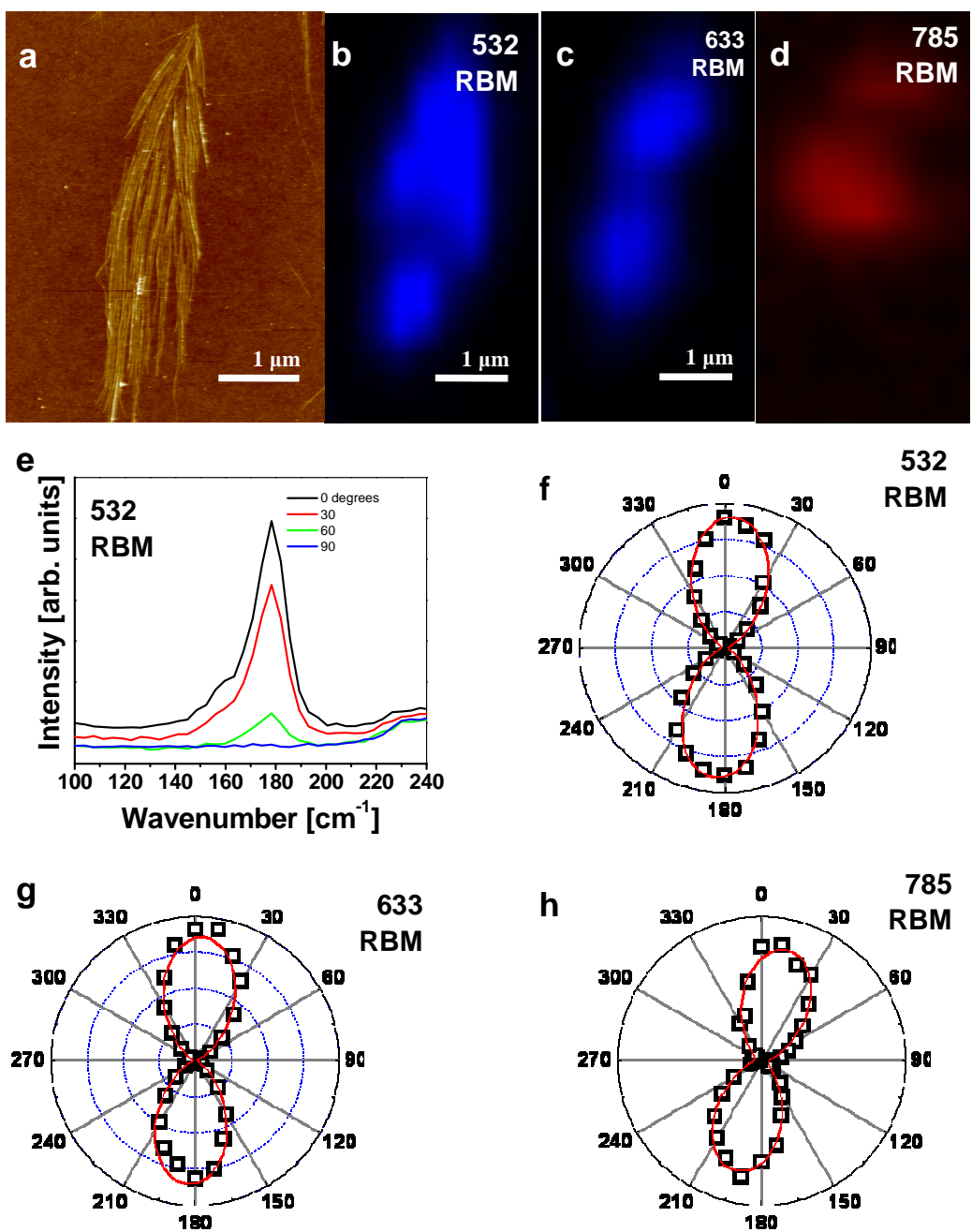

Figure 3

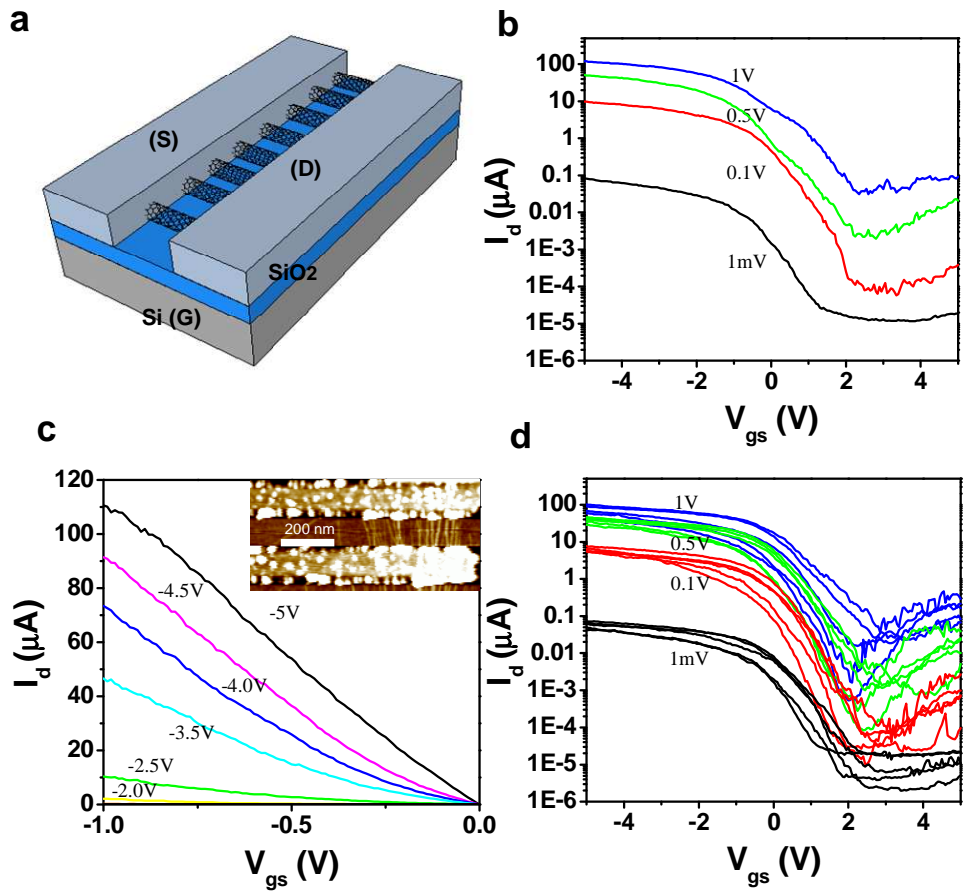

**Figure 4**

# Supplementary Information

# Self-Assembly of Semiconducting Single-Walled Carbon Nanotubes into Dense and Well-Aligned Rafts

Justin Wu[1,†], Liying Jiao[1,2,†], Alexander Antaris, Charina L. Choi[1], Liming Xie[1], Yingpeng Wu[3], Shuo Diao[1], Changxin Chen[1], Yongsheng Chen[3] and Hongjie Dai[1]*

1. Department of Chemistry and Laboratory for Advanced Materials, Stanford University, Stanford, California 94305, USA
2. Key Lab of Organic Optoelectronics & Molecular Engineering, Department of Chemistry, Tsinghua University, Beijing 100084, China
3. Key Laboratory of Functional Polymer Materials and Centre for Nanoscale Science and Technology, Institute of Polymer Chemistry, College of Chemistry, Nankai University, 300071, Tianjin, China

*†* These authors contributed equally to this work

*\* Correspondence to hdai@stanford.edu*

1. **Experimental details**
2. **Assignment of s- and m- SWNTs by Raman spectroscopy**
3. **Effect of surfactant composition and concentration**
4. **Effect of substrate surface modification and evaporation**
5. **Raman spectra of SWNTs rafts measured under three excitations**

## 1. Experimental details

**Separation of SWNTs by gel filtration**

20 mg SWNTs made by arc discharge[1] were dispersed in 10 mL 1 wt% SC aqueous solution by bath sonication for 1h, followed by centrifugation at 50,000 r.p.m. for 30 min to remove large aggregates and bundles. 4 mL of the supernatant was mixed with 12 mL 1 wt % SDS to obtain a ratio of SC/SDS=1/3 in the mixture while maintaining the total surfactant concentration at 1 wt%. The solution was added to a 1-cm-diameter column filled with 4 mL gel equilibrated with a solution of 0.25 wt% SC and 0.75 wt% SDS. It is known that SC disperses SWNTs more effectively than SDS at the same concentration. We gradually washed the column using four solutions with increasing percentages of SC and increasing 'dispersing power', 0.25 wt% SC/0.75 wt% SDS, 0.5 wt% SC/0.5 wt% SDS, 0.75 wt% SC/0.25 wt% SDS, and 1 wt% SC. The first two solutions with lower SC concentrations eluted mostly m-SWNTs and some unbound s-SWNTs indicated by the green to gray color of the eluted solution. The third wash with a 0.75 wt% SC/0.25 wt% SDS solution eluted a dark red fraction (S-fraction), indicating a high concentration of s-enriched SWNTs (Fig. 1c). The final washing with 1 wt% SC produced a fraction with a very faint pink color, suggesting this fraction contained little material.

**Separation of SWNTs by density gradient ultracentrifugation**

P2 SWCNTs synthesized by arc discharge (Carbon Solutions) were dispersed at a loading of 5 mg/mL in 7 mL of 1% SC using a horn sonicator at a tip power of 8 W for 1 hour immediately preceding density gradient ultracentrifugation. The dispersed SWCNTs were density shifted to 32% w/v iodixanol and to a final SDS:SC ratio of 1:4. Density gradient separations were performed in a homogenous 1% w/v surfactant loading of 1:4 SDS:SC throughout the entire centrifuge tube. The gradient architecture, from the bottom of the centrifuge tube, consisted of a 6.5 mL SWNCT solution (32% w/v iodixanol), a 15 mL linear density gradient (30–15 % w/v iodixanol), and an overlay (0% w/v iodixanol). All centrifuge tubes were run in a SW 32 Ti rotor (Beckman Coulter, Inc.) for 18 hours at 32,000 RPM and at a temperature of 22 °C. Following ultracentrifugation, fractions were collected at a vertical resolution of 0.5 mm using a piston gradient fractionator (Biocomp Instruments, Inc.)[4]. Due to the process differences, the CNTs were suspended in different surfactants than the gel filtered solution. The DGU sorted solutions were then dialyzed into 1% SC to remove the iodixanol and prepare them for raft making.

**Comparison of DGU and gel filtered solutions**

The DGU method provided SWNT solutions of higher semiconductor purity than the gel filtered solution, which is favorable for device applications. In order to judge which one should be used, individual devices were fabricated using the same processing steps and substrate as the raft devices using both the gel filtered as well as the DGU solution. The average current was found to be 3.3μA for the DGU solution, compared to 1.2μA from the gel filtered solution, indicating either more consistent high quality of the tubes or a slightly larger diameter distribution. From the on/off ratios of the devices, it was judged that the DGU solution had a better semiconductor purity of 98% compared to 97% from the gel filtered solution.

**Characterization of SWNT solutions**

UV-*vis*-NIR absorbance spectra of SWNTs solutions were measured by a Cary 6000i UV-*vis*-NIR spectrophotometer, background-corrected for any quartz contribution. The SWNT solution was drawn into capillary tubes for solution phase Raman spectroscopy. Raman spectroscopy measurements were taken using a Horiba HR800 Raman system with 532, 633, and 785 nm excitation. Solutions were spun onto $SiO_2$ for inspection via AFM using a Nanoscope IIIa multimode instrument in tapping mode. The mean SWNT length on the substrate was found to be ~ 0.7 μm, with tube lengths up to ~1.8 μm.

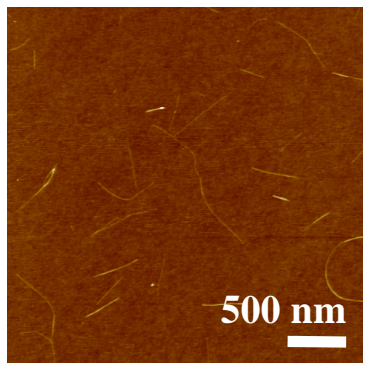

**Figure S1** AFM image of s-SWNT spun onto substrate.

**Assembly of SWNTs**

We first modified SiO$_2$/Si substrates with (3-aminopropyl) triethoxysilane (APTES) by placing a drop of APTES aqueous solution (1 mL in 5 mL H$_2$O) onto the substrate (~9 mm×6 mm), waiting for ~20 min, rinsing with water and blowing dry. After that, we placed a drop of solution (~60 μL) onto the APTES modified substrate, waited for ~1h at room temperature in atmosphere (the volume of droplet decreases by ca. a half at this point), blow-dried the substrate using directional air flow, heated the substrate at 80 $^{\circ}$C for two minutes, followed by a long rinse of the substrate with water for several hours to remove excess surfactants. For a 6×9 mm chip, the size of the drop used was 60 μL; the evaporation process required 40 minutes until the desired volume was achieved.

**Characterization of s-SWNTs rafts**

AFM images of rafts were taken with a Nanoscope IIIa multimode instrument in tapping mode. Polarized Raman was done on a Horiba HR800 Raman system with 532nm excitation. The laser power was kept at ~1 mW/μm$^2$ to avoid any heating effects. A half-wave plate was put in the laser path to rotate the polarization of the laser. An analyzer was put in the signal path to select the polarization of the Raman signal. Polarized Raman measurement was done in VV configuration, in which the polarization directions of the laser and the Raman signal were kept parallel and the Raman spectra were measured at different angles between the laser/Raman polarization direction and the raft direction. The number of SWNTs within a raft was counted by AFM; a feature of height of 1.2-1.7nm was counted as a single SWNT, height 1.7-2.2nm counted as a bundle of two SWNTs, and features of height 2.2-2.7nm counted as a bundle of 3.

**Fabrication of s-SWNTs rafts devices**

We used electron-beam lithography followed by electron-beam evaporation of palladium (20 nm) to fabricate a large array of source and drain electrodes on $SiO_2$/$p^{++}$ Si substrates with pre-deposited rafts. The devices were then annealed in Vacuum at 200 °C for 20 min to improve the contact quality.[2]

**2. Assignment of s- and m- SWNTs by Raman spectroscopy**

Semiconducting or metallic assignment for SWNTs is based on RBM frequency and the corresponding excitation energy in Kataura plot, in which the RBM-diameter relationship is $d=248$ cm$^{-1}$/$\omega_{RBM}$, where $d$ is diameter of SWNT and $\omega_{RBM}$ is RBM frequency.[3]

## 3. Effect of surfactant composition and concentration

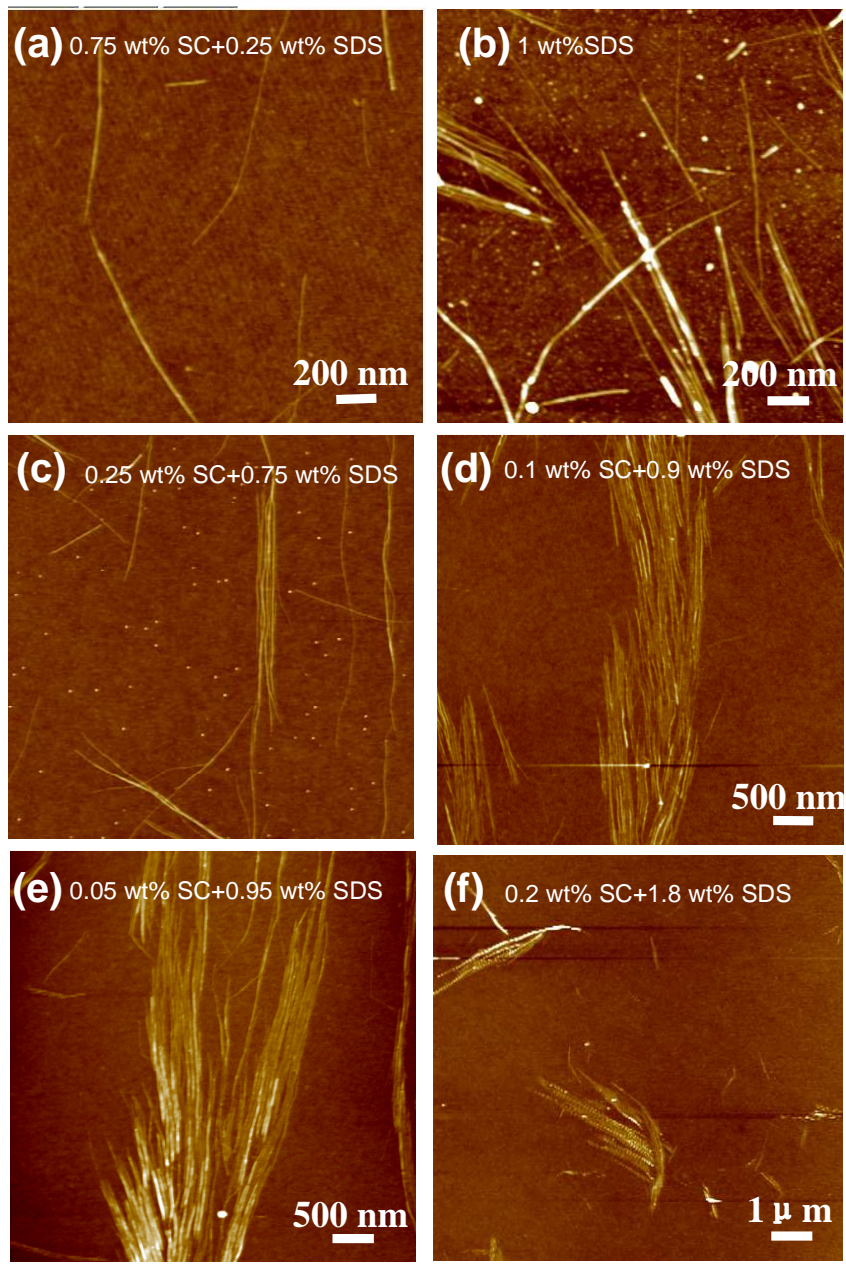

**Figure S2** AFM images of s-SWNT assemblies obtained at varied surfactant compositions: (a) 0.75 wt% SC+ 0.25 wt% SDS (S-fraction). Individual s-SWNTs randomly distributed on substrate. (b) 1 wt% SDS. Big bundles with some small assembly were observed. (c) (d) 0.1 wt% SC+ 0.9 wt% SDS (the optimized composition) (e) 0.05 wt% SC+ 0.95 wt% SDS. The thickness of rafts increased,

indicating multilayer assembly. (f) 0.2 wt% SC+ 1.8 wt% SDS. The ratio of SC/SDS was kept to 1:9 while the total concentration of surfactants increased from 1 wt% to 2 wt%. Big bundles appeared at this composition.

## 4. Effect of substrate surface modification and evaporation

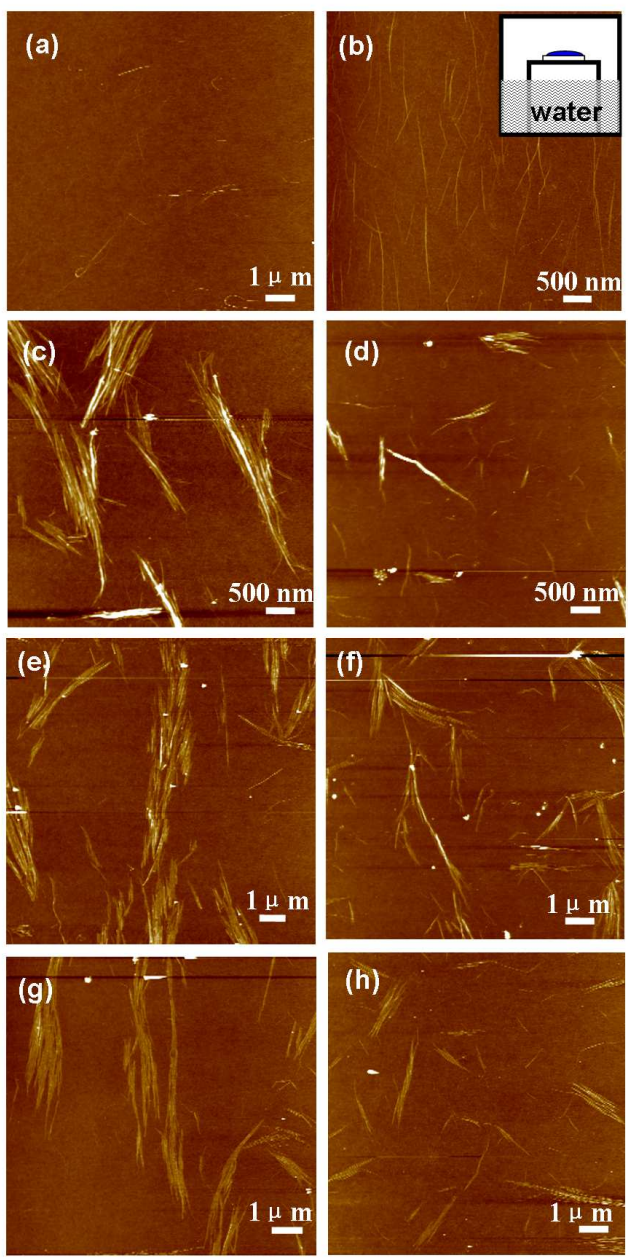

**Figure S3** (a) Deposition result of s-SWNTs on SiO$_2$/Si substrate without APTES modification. (b) s-SWNT assembly obtained in a sealed container with water. Inset, schematic of the set-up for slowing down water evaporation in the s-SWNT solution droplet. (c) and (d) s-SWNTs assemblies obtained after the volume of droplet decreased by 75% and 100%, respectively. Thicker rafts and big bundles were obtained after too much evaporation. ~50% was found to be the optimized volume decrease. (e) and (f) s-SWNTs assemblies obtained by evaporating ~50% water at 50 $^{o}$C and 60 $^{o}$C, respectively. Thicker rafts and big bundles were also obtained by too fast evaporation. Evaporation at 30 $^{o}$C to 40 $^{o}$C can produce flat rafts with aligned structures. (g) and (h) Rafts were obtained by removing remaining solution by directional blow-drying and wicking the solution away using a capillary tube, respectively. Obviously, directional blow-drying could roughly align the rafts.

**5. Raman spectra of SWNTs rafts measured under three excitations**

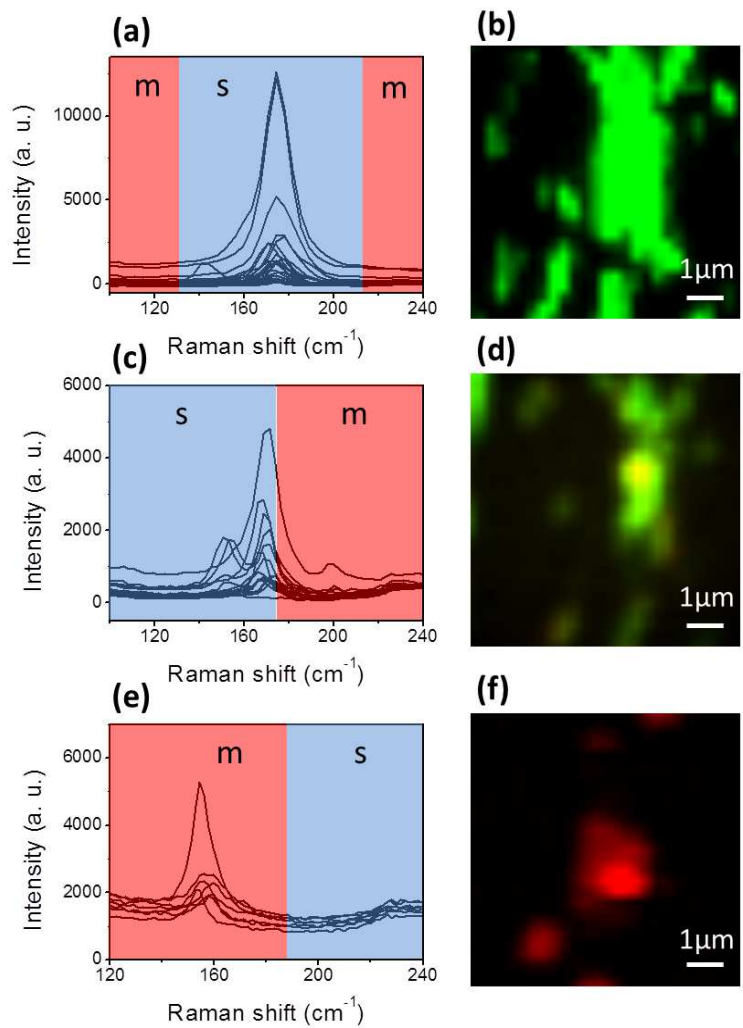

**Figure S4** Raman spectra of the s-SWNT raft shown in Fig. 3a under (a) 532 nm, (c) 633 nm, and (e) 785 nm excitation. Areas of graph covered in blue mark Raman shifts corresponding to s-SWNT RBM peaks. Areas in red mark Raman shifts corresponding to m-SWNT RBM peaks. The peak at ~225 cm$^{-1}$ is from the substrate. Raman mapping using (b) 532 nm, (d) 633 nm, and (f) 785 nm excitation. Areas containing s-SWNT RBM peaks are represented in green. Areas containing m-SWNT RBM peaks are shown in red